\documentclass{aa}
\usepackage{graphics}
\usepackage{times}
\begin{document}

\thesaurus{03           % Extragalactic astronomy
               (09.11.1;  % ISM: kinematics and dynamics
                11.09.1;  % Galaxies: individual
                11.09.4;  % Galaxies: ism
                11.11.1;  % Galaxies: kinematics and dynamics
                11.19.2)} % Galaxies: spiral

\title{Vertical motions in the disk of NGC 5668 as seen with optical 
Fabry-Perot 
spectroscopy\thanks{Based 
on observations made with the William Herschel Telescope operated on 
the island of La Palma by the Isaac Newton Group in the Spanish 
Observatorio del Roque de los Muchachos of the Instituto de
Astrof\'{\i}sica de Canarias}} 

\author{J. Jim\'enez-Vicente \inst{1}~ \inst{2}, E. Battaner \inst{2}}

\institute{Kapteyn Institute, Postbus 800, 9700 AV Groningen, The 
Netherlands
\and Departamento de F{\'\i}sica Te\'orica y del
 Cosmos, Universidad de Granada, E-18071, Granada, Spain}

\offprints{J. Jim\'enez-Vicente}
\mail{jjimenez@astro.rug.nl}
\titlerunning{Vertical motions in the disk of NGC 5668}
\date{Received October 21, 1999 / Accepted} 
\maketitle

\begin{abstract}
        We have observed the nearly face-on spiral galaxy NGC 5668 with 
the TAURUS II 
Fabry-Perot interferometer at the William Herschel Telescope using the 
$H\alpha$ line
to study the  kinematics of the ionized gas. From the extracted 
data cube we construct 
intensity, velocity and velocity dispersion maps. We calculate the 
rotation curve in the 
innermost 2 arcmin and we use the residual velocity field to look for 
regions with 
important vertical  motions. By comparing the geometry of these 
regions in the residual 
velocity field with the geometry in the intensity and velocity 
dispersion maps we are 
able
to select some regions which are very likely to be shells or chimneys 
in the disk. The 
geometry and size of these regions are very similar to the shells or 
chimneys detected in 
other galaxies by different means. Moreover, it is worth noting than 
this galaxy has
been reported to have a population of neutral hydrogen high velocity 
clouds (Schulman 
et al. 1996) which, according to these observations, could have been 
originated by 
chimneys similar to those reported in this paper.

\end{abstract}

\section{Introduction}

        There is now great observational evidence that  disk-halo 
interactions in 
galaxies as well as the structure of the interstellar medium (ISM) is 
closely related to
star formation processes in the disks of spiral galaxies 
(see the review by Dahlem 1997).
 Big shells develop around the brightest star forming regions, induced 
by the energy 
input of supernovae and the strong stellar winds produced by high-mass 
stars. 
These shells
 can grow enough to be able 
to break the disk, allowing large amounts of gas to  blow out from the 
disk along these
big chimneys (Norman \& Ikeuchi 1989). The very hot gas going out 
through these chimneys cools as it rises 
until it eventually 
recombines and condenses to form clouds of neutral gas that 
fall back to the 
plane (Shapiro \& Field 1976, Bregman 1980).
This {\em fountain} model then provides an explanation for the 
origin of high velocity clouds (HVC's) that
have been observed in our galaxy and a in few external galaxies (with the 
galaxy studied in 
this paper being one of those few (Schulman et al. 1996, hereafter S96)), 
although 
alternative
explanations have been proposed as well (Blitz et al. 1999). 
Good reviews on
this topic can be found in Wakker \& van Woerden (1997) and 
van der Hulst (1996, 1997). On the other 
side, the expanding shells can induce new star formation 
(sequential star formation (SSF)) at their 
edges. This effect has already been observed in some
galaxies (see for example Thilker et al. 1998, hereafter T98) 
and it is also observed 
to take place in the case of NGC 5668 in our observations. 

        According to the chimney model, the structure of the ISM, and 
in particular 
whether the chimney phenomenon takes place  or not, is controlled by 
the amount of 
star formation. The study of the properties of these phenomena in a 
sample of nearby 
galaxies can greatly help to understand the structure of the ISM and 
the nature of
disk-halo interactions. Observations of the neutral gas and narrow band 
imaging
of the ionized gas have already been extensively used to study these 
phenomena
(see the review by Dahlem 1997). Scanning long-slit
  $\mathrm{H\alpha}$ spectroscopy has also been used to study these
phenomena (Saito et al., 1992; Tomita et al., 1993, 1994). In this 
work we make a first attempt to use optical Fabry-Perot spectroscopy in 
a nearly 
face-on spiral galaxy to directly study these vertical motions. 
Therefore we
have chosen the spiral galaxy NGC 5668 which is already known to have 
HVC's and
an important rate of star formation. These facts make it a perfect 
candidate for us
to detect important vertical motions in its disk.

In Sect. 2 we describe the general properties of the galaxy NGC 5668, 
with particular
emphasis on the observation of the HVC's. Sect. 3 deals with the 
observations and
data reduction (including calculation of intensity, velocity and 
velocity 
dispersion maps).
Sect. 4 is devoted to the calculation of the rotation model for the 
galaxy from the
observed velocity field. Sect. 5 describes how the 
residual velocity field is used
to look for systematic deviations of circular rotation and how 
comparison of
the geometry of the residual velocity field with that of the intensity 
and velocity 
dispersion in some regions can be used to detect real shells and/or 
chimneys. 
Finally, in 
Sect. 6 we report the shell candidates found in NGC 5668 and some of 
their properties. 

\section{NGC 5668: General properties}

\begin{figure}[tbp]
\resizebox{8.8cm}{!}{\includegraphics{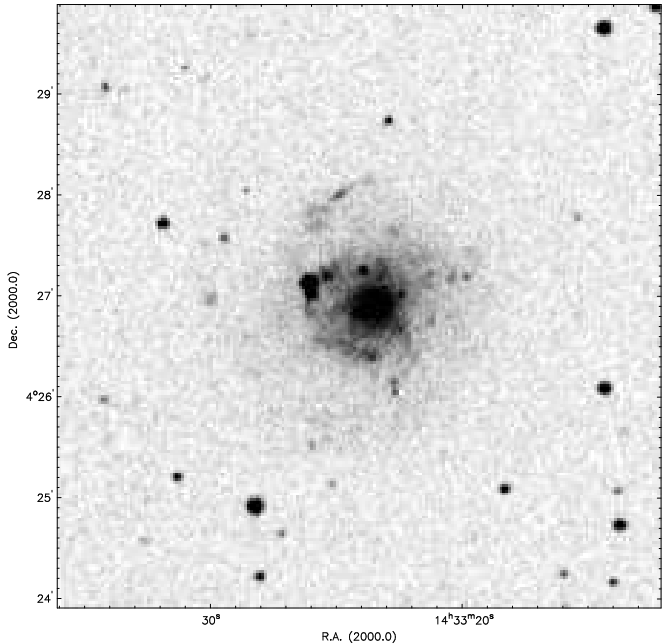}}
\caption{Optical image of NGC 5668}
\label{fig:dss}
\end{figure}

        NGC 5668 is a nearly face-on (inclination $ \approx 18^\circ $) 
late type 
spiral galaxy (Sc(s) II-III). A wide band optical image of the galaxy 
can be seen in
Fig. \ref{fig:dss}\footnote{
Based on photographic data of the National Geographic Society -- Palomar
Observatory Sky Survey (NGS-POSS) obtained using the Oschin Telescope on
Palomar Mountain.  The NGS-POSS was funded by a grant from the National
Geographic Society to the California Institute of Technology.  The
plates were processed into the present compressed digital form with
their permission.  The Digitized Sky Survey was produced at the Space
Telescope Science Institute under US Government grant NAG
W-2166.}. We will adopt a distance of 22.6 Mpc, consistent with 
$H_0=70~ \mathrm{km~ s^{-1}~Mpc^{-1}}$. Its size (the blue isophote at
25 mag/arcsec is 3.3 arcmin in diameter) makes it an ideal target for
observation with TAURUS II (which has a FOV of about 5 arcmin)).
The measured radial scale length for NGC 5668 is 27.3 arcsec, which 
means
that our observations reach up to around 4.4 scale lengths.
The FIR and $\mathrm{H\alpha}$ luminosities of NGC 5668 (S96)
show that it has an important amount of star formation.
A summary of its main properties is shown in Table \ref{tab:prop}.

\begin{table}
\caption{Galaxy properties}
  \label{tab:prop}
  \begin{tabular}{ll}
    \hline \\
    Name             & NGC 5668 \\
    Type             & Sc(s) II-III \\
    R.A.(2000)       & $14^h33^m24.9^s$ \\
    Dec (2000)       & $4^\circ 27'02''$ \\
    B Magnitude      & 12.2 \\
    $D_{25}$         & $3.3'$ \\
    $V_\odot$ (RC3)  & 1583 km/s \\
    Distance         & 22.6 Mpc \\
    Inclination      & $18^\circ$ \\
    P.A.             & $145^\circ$ \\
%    Environment      & Ursa Major Cluster. \\
%                     & No close ($<100$kpc) companions \\
    \hline
   \end{tabular}
\end{table}

NGC 5668 was observed in the HI line by means of the Arecibo telescope 
by
Schulman et al. (1994). They detected high velocity wings in the line 
shape, 
which were attributed to HVC's in the galaxy. As the Arecibo observations
did not contain information on the spatial distribution of gas, 
subsequent 
observations were taken with the VLA telescope by S96.
These observations confirmed the existence of the HVC's. They found
that the amount of mass on neutral hydrogen in HVC's is about 
$4\times10^8~M_\odot$. Their observations lacked sufficient spatial 
resolution (the FWHM of the synthesized beam was $48''\times 41''$) to
detect the shells or chimneys in the disk that could be the origin of
this large amount of gas in high velocity components. The main purpose
of this paper is to try to detect such features by means of
Fabry-Perot spectroscopy.

\section{Observations and data reduction}

        The observations were taken on March 19, 1997, at the 
William Herschel Telescope using the TAURUS II Fabry-Perot 
interferometer 
at the Cassegrain focus, with
the f/2 camera and the 500$\mu$ etalon. The detector used was a TEK CCD.
With this configuration the pixel size is 0.56 arcsec. We rebinned 2x2 
the
CCD reading, resulting in a pixel size of 1.12 arcsec (which for an 
adopted
distance to NGC 5668 of 22.6 Mpc results in 123 pc/pixel). The observing
conditions were not photometric (therefore we were not able to make
photometric calibrations and we will use arbitrary units for intensity
throughout this paper). The measured seeing in the final images was 3 
arcsec
which is still a very good spatial resolution when compared with 
the previously mentioned kinematical data available for this galaxy. 
We used the $\mathrm{H\alpha}$ line to trace the distribution and
kinematics of the ionized gas. According to the observed velocity of the
galaxy (see Table \ref{tab:prop}) the wavelength for the redshifted 
 $\mathrm{H\alpha}$ 
line is 6597.5 
$\mathrm{ \AA}$. Therefore we used 
the 
6601/15 filter for order sorting. We scanned the free spectral range 
(FSR) in
55 steps with an exposure time of 120 seconds per frame. This gave a
spectral resolution of 3.61 km/s/pixel. Calibration datacubes 
illuminating
the instrument with a CuNe lamp were taken at the beginning and at the 
end
of the night. Moreover a ring calibration frame was taken before and
after the datacube exposure to make the RVT correction. 
The instrumental width (measured after 
phase correcting the calibration datacube) was 7.7 km/s. 
The main properties of the observational
setup are summarized in Table \ref{tab:setup}.

\begin{table}
\caption{Observational parameters}
  \label{tab:setup}
  \begin{tabular}{ll}
    \hline \\
    Date of observation & 19/03/1997 \\
    Telescope  & WHT \\
    Focus      & Cassegrain \\
    Instrument & TAURUS II \\
    Etalon     & $500 \mu$ \\
    Filter     & 6601/15 \\
    Detector   & TEK CCD \\
    Steps      & 55 \\
    Exposure  time per step  & 120 sec \\
    Free Spectral Range & 194.1 km/sec \\
    Instrumental width & 7.7 km/s \\
    Spectral resolution  & 3.61 km/sec/pixel \\
    Pixel size   &  1.12 arcsec \\
    Seeing       &  3 arcsec \\
    \hline
   \end{tabular}
\end{table}

\begin{figure}[tbp]
\resizebox{8.8cm}{!}{\includegraphics{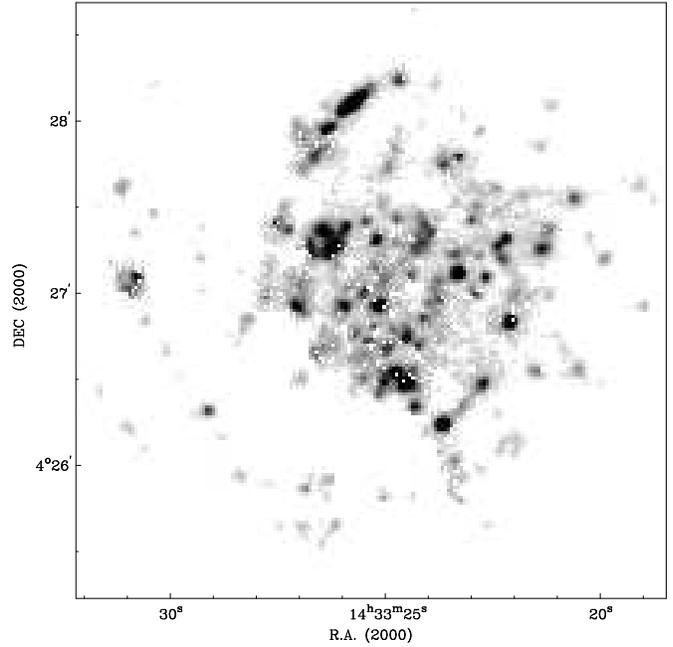}}
\caption{Observed $\mathrm{H\alpha}$ intensity map for NGC 5668}
\label{fig:inten}
\end{figure}

        After phase and wavelength calibration the datacube was 
bias subtracted. The continuum level was calculated by fitting a 
first order polynomial to the
channels where there is no line emission from the galaxy, and the 
continuum map was then subtracted from the datacube.
From the continuum free datacube we calculated the intensity, 
velocity and velocity dispersion maps. We calculated the maps by 
two different methods: a moments procedure and a gaussian fitting. 
The intensity and velocity maps are very similar in both cases 
and we use the ones calculated by the gaussian
fitting procedure. The velocity dispersion maps are 
substantially different and we use the one calculated by 
the fitting procedure as well, in order
to avoid systematic bias as pointed out by van der Kruit \& Shostak 
(1982). The calculated maps are very noisy. To clean these maps we 
imposed certain conditions to flag out bad data points. 
We therefore kept only those pixels which had a velocity 
dispersion between 7.7 km/s (the instrumental width) and
50 km/s (data points above this level are clearly not valid). 
We also required the points to have a peak intensity at least 2 
times larger than the noise 
dispersion. Finally some clearly bad data points were excluded 
interactively by inspection
of the resulting maps. As we are interested in the non-thermal
velocity dispersion, we corrected the calculated dispersion for
the natural width of the line (3 km/s), the insrtumental width 
(7.7 km/s) and the thermal width (9.1 km/s assuming a temperature
of $10^4$ K for the ionized gas). The corrected 
average velocity dispersion calculated for this galaxy is
around 16.5 km/s independent of radius. This value is somewhat
higher than that measured for other galaxies, which is usually
around 10 km/s (see for example Jim\'enez-Vicente et al. 1999),
showing that the ISM of NGC 5668 is more {\em turbulent}.
The $\mathrm{H\alpha}$ intensity map for 
NGC 5668 is shown in Fig. \ref{fig:inten}.

\section{The velocity field and the rotation curve}

\begin{figure}
\resizebox{8.8cm}{!}{\includegraphics{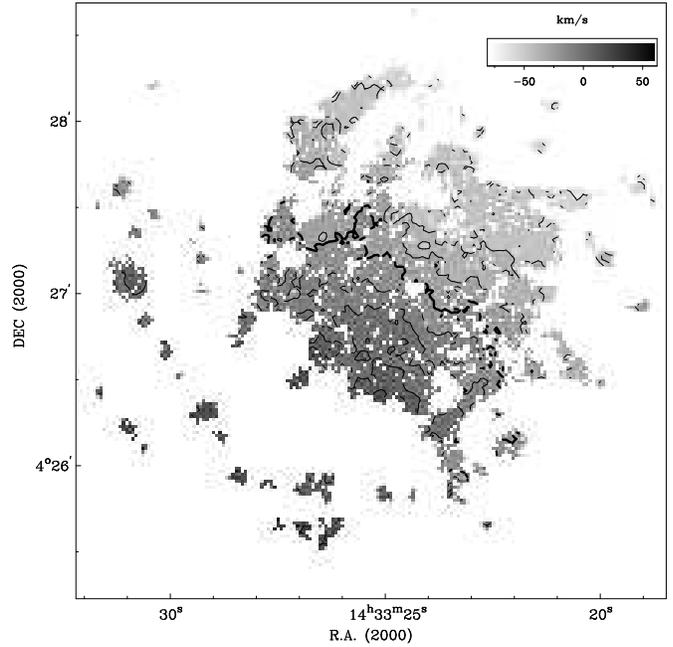}}
\caption{Observed velocity field for NGC 5668. Contour levels are shown
each 10 km/s with the thick line corresponding to 0 km/s}
\label{fig:vel}
\end{figure}

\begin{figure}
\resizebox{8.8cm}{!}{\includegraphics{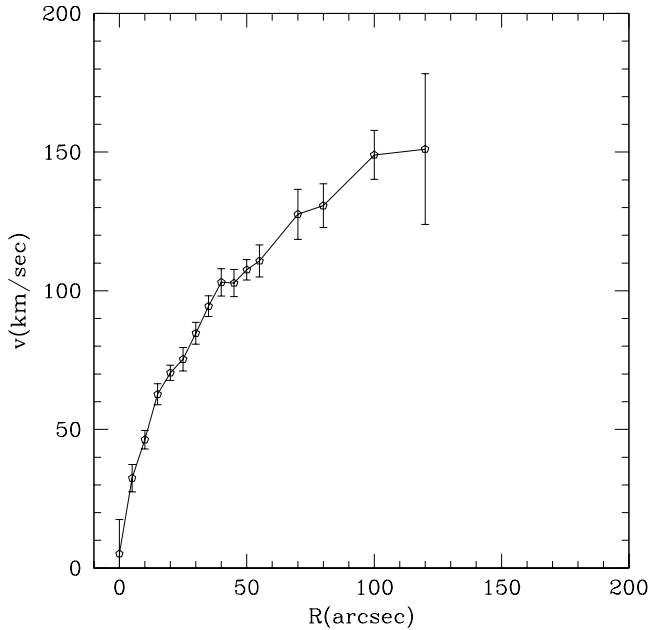}}
\caption{Calculated rotation curve for NGC 5668. Error bars represent
least-squares fit errors.}
\label{fig:rot}
\end{figure}

        The observed velocity field is shown in Fig. \ref{fig:vel}.
The velocity field is well mapped out to 1.5 arcmin from the centre. 
Outside this radius only a few H II regions are visible and therefore 
the
extracted kinematic information for this region is less reliable.
In order to calculate a rotation model for the galaxy, we used the
standard procedure of fitting a tilted rings model (see Begeman 1989) 
to
the observed velocity field. In order to minimize the number of free 
parameters and to simplify the fitting process we used known information
from previous observations.
Therefore we fixed the position of the kinematic centre at the
same position than the optical centre. We fixed the inclination and 
position angles at $18^\circ$ and $145^\circ$ respectively, according
to the values reported
by S96, which allow us to make a better 
comparison with their results. These authors have shown that both
angles remain constant in the inner 2 arcmin, which is the region
for which we have data, and we therefore know that our results are
not affected by this assumption. The systemic velocity for the galaxy
was also taken from S96 to be 1582 km/s. 
We have checked that small variations of these parameters do not 
substantially change the model velocity field. The 
only free parameter left per ring, namely the rotational velocity, 
is fitted to the observed velocity field. Finally, to take
into account that we have fewer data points in the outer part of the
galaxy, we make the rings wider in this region to ensure that
a minimum number of data points is included in a ring. This results
in the outer rings
being 4 times wider than the inner ones. The resulting 
rotation
curve is shown in Fig. \ref{fig:rot}. It 
rises continuously out to about 100 arcsec from the centre, where it 
seems to become flat at a velocity of about 150 km/s. 
This value is somewhat higher
than the rotational velocity reported by S96 of 130 
km/s. In fact, the rotation velocities determined at every radius 
are always higher than
those previously reported by S96 by around 20 km/s.
It is worth noting than the last two data 
points in this curve
are not fully reliable, as we remarked above that we only have
a few H II regions lying outside 1.5 arcmin from the centre (most of
them in the receding part of the galaxy) from which
kinematical information can be obtained. If a value of
130 km/s is adopted for the rotational velocity in the 
outer part of the galaxy to match the observations in HI, then all 
the outermost regions, which trace one of the spiral arms, would 
have 
quite high 
residual velocities. Although this is not unlikely, 
we have no way to confirm that this is what
is happening, and we therefore prefer to use
the best fit to our data to calculate the residual velocity field and thus
we use
the rotation curve shown in Fig. \ref{fig:rot}. In the inner
1.5 arcmin, where our determination of the  rotation curve is reliable, 
this
is the best determination of the rotation curve (in both spatial and 
spectral
resolution) for NGC 5668 up to date.

\section{The residual velocity field and the shell/chimney candidates 
(HRVR's)}

\begin{figure}
\resizebox{8.8cm}{!}{\includegraphics{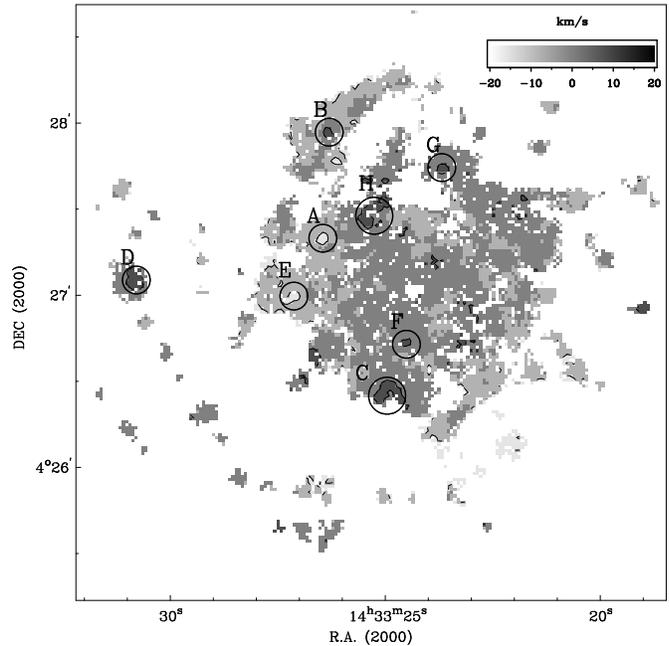}}
\caption{Residual velocity map for NGC 5668. The contours are for
residuals higher than 10 km/s. The circles and letters mark the
HRVR's (see text).}
\label{fig:res}
\end{figure}

        The residual velocity field is calculated by subtracting the 
rotation model from the observed velocity field. Fig. \ref{fig:res}
shows the resulting map. 
Although the map is quite noisy, a close 
inspection
allows us to detect some regions which have a systematic deviation 
from
rotational velocity greater than 10 km/s (which is 1.5 times the
dispersion of the residual velocity map). We call these 
regions {\em high residual velocity regions} (HRVR's). In order to
distinguish between HRVR's with real important vertical motions 
from regions whose residual could be
an effect of noise or random motions in the disk, we
also use the size and shape of the region. We therefore
label as a HRVR only to those regions with a clear symmetry
in the shape and with a mean diameter between 3 (which is our
spatial resolution) and 13 arcsec (which corresponds to
about 1.5 kpc). 
The regions have been marked with a circle and named with a letter
in Fig. \ref{fig:res}. 
The shape of the HRVR's
(which is in many cases nearly circular) makes it very unlikely 
that those regions are related  to streaming motions. 
It then seems logical to interpret the residuals as true 
vertical motions in the disk.
This interpretation is strongly reinforced if we look at the intensity
map in the locations where the HRVR's are found. We can see that
the regions are clearly associated with regions of star formation which
fall in their centres or immediate surroundings. In most cases
the structure of the velocity dispersion map in the HRVR's is also
strongly correlated with both intensity and residual velocity (reaching
values up to 34 km/s in the centre of the HRVR), which
supports the idea of regions of great activity (high temperature
or violent motions). These facts strongly
support the hypothesis that the HRVR's are regions with real vertical
motions of the ionized gas which are related to star forming processes
in the disk. The most straightforward interpretation of these features
(although certainly not the only one)
is that they are shells or chimneys (depending on their size and age
and on whether they have been able to {\em break} the disk and to
blow gas out of it)
formed around highly active star-forming regions by the strong stellar
winds or correlated SN explosions of the multiple massive young stars
found whithin those regions. 
This scenario perfectly matches the chimney
model
proposed by Norman and Ikeuchi (1989) for the ISM and is also supported
by observations in other galaxies (see for example T98).
The fact that NGC 5668 has a great amount of star formation
and that HVC's of neutral hydrogen have been detected in it 
(S96) makes the whole scenario fully coherent.

\section{Some shell/chimney properties}

\begin{figure}
\resizebox{8.8cm}{!}{\includegraphics{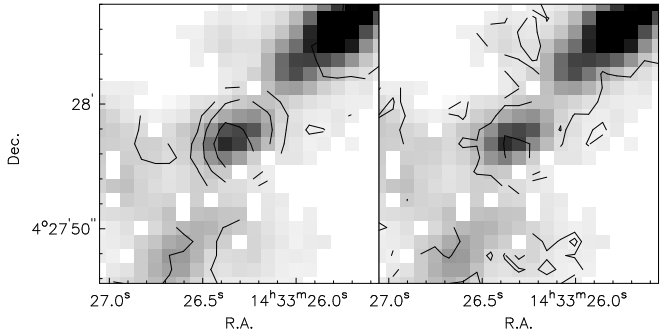}}
\caption{Example of a {\em circular} region: Region B. 
Left panel shows a greyscale map of
$\mathrm{H\alpha}$ intensity with the residual velocity contours (from 
0 (outer) to 15 (inner) km/s
in steps of 5 km/s). Right panel
shows a greyscale map of $\mathrm{H\alpha}$ intensity with the 
velocity dispersion contours (13 (inner) and 18 (outer) km/s in steps
of 5 km/s).}
\label{fig:mosb}
\end{figure}

\begin{figure}
\resizebox{8.8cm}{!}{\includegraphics{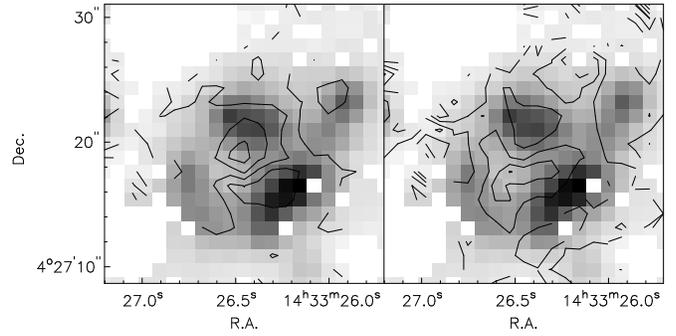}}
\caption{Example of a ring-like region: Region A. 
Left panel shows a greyscale map of
$\mathrm{H\alpha}$ intensity with the residual velocity contours 
(from -15 (inner) to 0 (outer) km/s
in steps of 5 km/s). Right panel
shows a greyscale map of $\mathrm{H\alpha}$ intensity with the 
velocity dispersion contours (from 15 (outer) to 30 (inner) km/s 
in steps of 5 km/s).}
\label{fig:mosa}
\end{figure}

\begin{figure}
\resizebox{8.8cm}{!}{\includegraphics{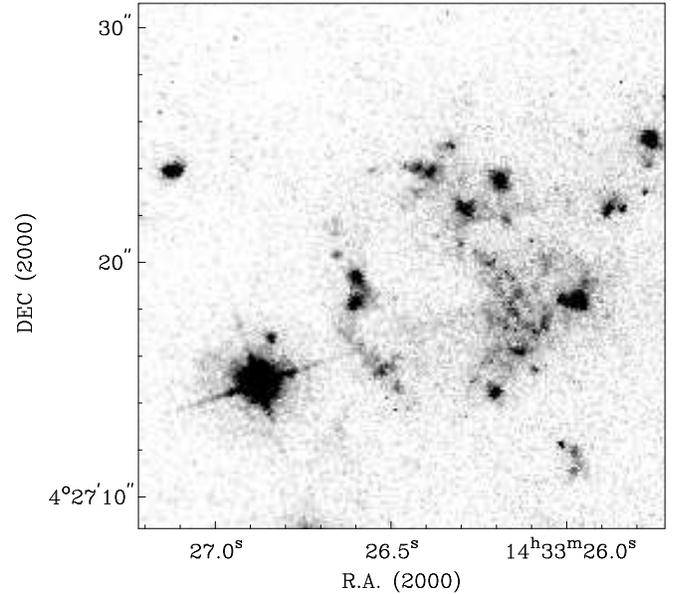}}
\caption{Region A as seen by the WFPC/2 of the HST with the F658N
  filter.}
\label{fig:hst}
\end{figure}

\begin{table}[ht]
\caption{Shell/chimney properties. Columns show 1) The HRVR's ID according
to Fig. \ref{fig:res}, 2) Estimated radius in pc, 3) Maximum re\-si\-dual
velocity, 4) Average velocity dispersion around region centre, 
5) Type: R=Ring-like, C=circular,
U=undefined.}
{\normalsize
  \label{tab:prop2}
  \begin{tabular}{|c|c|c|c|c|}
    \hline
    ID & R (pc)& $\mathrm{V_{max}}$ (km/s)& 
$\mathrm{<\sigma>}$ (km/s) & Type \\ \hline
A & 500 & -17 & 18.5 & R \\
B & 200 & 13  & 17.7 & C \\
C & 600 & 13  & 19.5 & R \\
D & 300 & 15  & 19.6 & C \\
E & 400 & -18 & 22.4 & U \\
F & 600 & 12  & 17.0 & R \\
G & 250 & 13 & 16.0 & C \\
H & 450 & 15 & 17.1 & U \\
    \hline
   \end{tabular}
}
\end{table}

\begin{figure}
\resizebox{8.8cm}{!}{\includegraphics{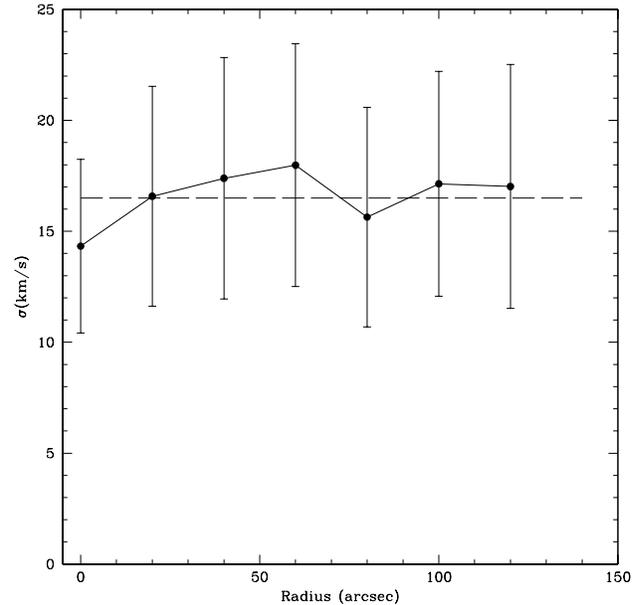}}
\caption{Radial distribution of the corrected velocity dispersion for 
NGC 5668. Error bars represent the dispersion within a ring. 
The dashed horizontal line represents the average value
of 16.5 km/s.}
\label{fig:dvr}
\end{figure}

        We were able to find eight HRVR's in NGC 5668. When we look
at the positions of the HRVR's in the intensity and velocity dispersion
map, we find that they are always related to regions of active
star formation.
Whether the star formation regions
lie in the centre or the surroundings of the HRVR's can probably
be explained as an age effect. The youngest, yet {\em small} 
expanding shells 
show the star-forming regions at their centre and
are therefore roughly {\em circular} in the intensity map. Fig. 
\ref{fig:mosb} shows an example of this type. 
The more evolved stalled chimneys are surrounded by a more or less
complete
ring-like structure of star-forming
regions which are probably formed by sequential star formation
(see T98). Fig. \ref{fig:mosa} shows one of
these cases.    
Both Figs. show clearly that there is a strong correlation
between residual velocity, intensity and velocity dispersion in
the HRVR's. In the case of region B (Fig. \ref{fig:mosb}) the
peak in the residual velocity perfectly matches the position of
a bright HII region and of a minimum in the velocity dispersion.
In the case of region A (Fig. \ref{fig:mosa}) the
residual velocity peaks roughly at the centre of a ring-like
structure formed by HII regions, and
the velocity dispersion peaks right in the centre of this ring-like
structure in the low intensity part. 
The structure of the $H\alpha$ 
emission around region A can be seen better in fig. \ref{fig:hst} 
where this region, extracted from the image of NGC 5668 taken
with the WFPC/2 of the HST\footnote{Based on observations made with 
the NASA/ESA Hubble Space Telescope, obtained from the data archive at
    the Space Telescope Science Institute. STScI is operated by 
the Association of Universities for Research in
    Astronomy, Inc. under NASA contract NAS 5-26555} using the F658N
filter is shown. This filter perfectly matches the wavelength of
the $H\alpha$ line for NGC 5668, and although it also contains the 
NII line and continuum emision, it mostly traces the ionized gas.
Fig. \ref{fig:hst} shows that there are not only compact HII
regions, but also a large amount of diffuse gas emission around this
region. This can
also be seen in most of the other HRVR's detected in NGC 5668 which are
visible in the HST image. 
%Only two exposures of 700 sec are available in the archive, which 
%makes the removal of all the cosmic ray events on the chip almost
%imposible.
Although the structure of
most of the HRVR's do not show this aspect (for example many of the
ring-like structures are not as complete as the shown example, 
but show just half a ring or even less), there is always a 
clear correlation between the structure of residual velocity,
intensity and velocity dispersion. Moreover the 
HST image shows that around these HRVR's (seven of the eight detected
are visible in the HST image) there are not only
numerous HII regions, but also a large amount of difuse ionized gas.
Table \ref{tab:prop2} gives some of the properties of the
shell/chimneys detected in NGC 5668. It is worth noting that all the 
HRVR's (except region G) have an average velocity dispersion above the
average of the galaxy (which is around 16.5 km/s throughout the 
whole disk as shown in Fig. \ref{fig:dvr}). As we noted previously, 
the HRVR's are always in or surrounded by regions of
high velocity dispersion. 

Although it is difficult to make
an estimation for the ages of the shells, previous studies
(T98) show that they range from a few Myr for
the small expanding shells up to some tens of Myr for the largest
stalled chimneys.

\section{Conclusions}

        We have carefully analyzed the data obtained for NGC 5668 with
the Fabry-Perot interferometer TAURUS II at the WHT in order seek
a connection between the star formation processes and vertical
motions in spiral galaxies. We have found that there is a clear 
correlation between the morphology of the regions with a high residual
velocity (HRVR's) and the intensity of the $\mathrm{H\alpha}$
 emission, showing
that the HRVR's are indeed regions with important vertical motions
associated with star formation processes. Although we are not able
to calculate the ages and energetics of these features, comparison
with observations in other galaxies strongly supports the hypothesis
that the structures detected present a wide age range, from
young expanding shells in a bright HII region, to evolved chimneys
blowing out hot gas to the halo, surrounded by several
bright HII regions. The formation of these regions was probably
induced by the pressure exerted by the expanding shell/chimney on
the ambient gas. An alternative explanation to these features is that
they are produced by infall/collision of gas clouds with the disk
(e.g. Saito et al., 1992) followed by induced active star
formation. Although this scenario explains in a very
natural and simple way the fact that the velocity
offsets are one-sided, this is also quite normal in expanding shells.
If they are formed not exactly in the equator but slightly off-plane, 
they will grow up mostly in the low density side, therefore
showing only one-sided offsets in the velocity structure. In fact, with
the present data there is no way to decide whether the HRVR's are
moving into or out of the disk. On the other hand, the different 
structures found in the HRVR's (from compact HII regions to {\em
rings} of HII
regions) are fully compatible with an evolutionary pattern in the 
chimney model.
Therefore, although we can not rule out the 
infalling hipotheses with the present
observations, we find the chimney scenario followed by sequential
star formation in 
the shell borders to be a more likely explanation for what is 
happening in NGC 5668.
These observations provide in any case a clear link between the
star formation processes in the disk with other observed phenomena
in NGC 5668 like high HVC's of neutral hydrogen.
Observations of this kind have been made for other galaxies 
with a lower star formation rate and without HVC's, and the residual
velocity field is not reported to show these features (see for example 
Jim\'enez-Vicente et al. 1999, for a similar study of NGC 3938).
This fact seems to suggest that these features and
the existence of HVC's in the disk are
closely related. High resolution HI observations of NGC 5668 would
be desirable to confirm the connection between those phenomena.

\end{document}